\documentclass{svjour3}                     
\usepackage{graphicx}
\usepackage{soul}
\usepackage{color}
\usepackage{url}
\usepackage{tikz}
\usepackage{natbib}
\setcitestyle{authoryear,open={(},close={)}}

\usetikzlibrary{shapes.geometric, arrows, fit}
\tikzstyle{io} = [ellipse, text width=3cm, minimum height=1cm, text centered, draw=black]
\tikzstyle{process} = [rectangle, minimum width=3cm, minimum height=1cm, text centered, text width=3.8cm, draw=black]
\tikzstyle{decision} = [diamond, text width=2cm, text centered, draw=black]
\tikzstyle{arrow} = [thick,->,>=stealth]
\tikzstyle{arrow2} = [thick,<->,>=stealth]

\begin{document}

\title{EmoconLite: Bridging the Gap Between Emotiv and Play for Children With Severe Disabilities}
\titlerunning{Emoconlite}

\author{Javad Rahimipour Anaraki, Chelsea Anne Rauh, Jason Leung, Tom Chau}

\institute{J. R. Anaraki,  J. Leung and T. Chau\at 
        Institute of Biomedical Engineering, University of Toronto, Toronto, ON, Canada\\
        \email{j.rahimipour@utoronto.ca, jasonwh.leung@mail.utoronto.ca, \& tom.chau@utoronto.ca}
        \and
       J. R. Anaraki, C. A. Rauh,  J. Leung and T. Chau\at
        Holland Bloorview Kids Rehabilitation Hospital, Toronto, ON, Canada\\
        \email{crauh@hollandbloorview.ca}
}

\date{Received: date / Accepted: date}

\maketitle

\begin{abstract}
Brain-computer interfaces (BCIs) allow users to control computer applications by modulating their brain activity. Since BCIs rely solely on brain activity, they have enormous potential as an alternative access method for engaging children with severe disabilities and/or medical complexities in therapeutic recreation and leisure. In particular, one commercially available BCI platform is the Emotiv EPOC headset, which is a portable and affordable electroencephalography (EEG) device. Combined with the EmotivBCI software, the Emotiv system can generate a model to discern between different mental tasks based on the user's EEG signals in real-time. While the Emotiv system shows promise for use by the pediatric population in the setting of a BCI clinic, it lacks integrated support that allows users to directly control computer applications using the generated classification output. To achieve this, users would have to create their own program, which can be challenging for those who may not be technologically inclined. To address this gap, we developed a freely available and user-friendly BCI software application called EmoconLite. Using the classification output from EmotivBCI, EmoconLite allows users to play YouTube video clips and a variety of video games from multiple platforms, ultimately creating an end-to-end solution for users. Through its deployment in the Holland Bloorview Kids Rehabilitation Hospital's BCI clinic, EmoconLite is bridging the gap between research and clinical practice, providing children with access to BCI technology and supporting BCI-enabled play.

\end{abstract}

\section{Introduction}
\label{intro}
Brain-computer interfaces (BCIs) allow users to interact with computers using only their brain activity. By performing mental tasks that generate specific brain activity patterns, users can use BCIs to control various applications and devices \citep{mak2009clinical}. When used by children with severe disabilities and/or medical complexities in the context of play and leisure, BCIs may open up a world of possibilities by reducing leisure disparities and supporting active, independent and engaged participation in leisure and life. However, one of the barriers preventing the widespread use of BCIs is their accessibility outside of the research environment. To this end, a program that facilitates access to BCI at home and in the clinic can bridge the gap between research and clinical practice.

Through collaboration with children with severe disabilities and/or medical complexities, and their families, a BCI clinic can support the well-being and quality of life of the pediatric population on an ethically sustainable research foundation, ultimately contributing to the advancement of BCI technology and further benefit for its users \citep{mietola2017voiceless}. Besides, such a program can also facilitate an environment of fun, function, fitness, family, friendship and play that forms the foundation of sustained skill development and a holistic perspective to healthcare \citep{rosenbaum2012f}. As an extension to the BCI clinic a therapeutic recreation (TR) BCI program was created to facilitate participation and continued practice of BCI skills while fostering opportunities for socialization. As one of the founding partners of BCI Canada collaborative network (BCI-CAN), Holland Bloorview Kids Rehabilitation Hospital has taken on this mission through its BCI clinic, which introduces children with severe disabilities and/or medical complexities, and their families to BCIs \citep{kinney2020advancing}. 

As the ease of access to mobile technologies was a prerequisite for iPads and smart phones to impact the field of augmentative and alternative communication \citep{mcnaughton2013ipad}, we expect that the accessibility of BCIs will be a critical factor contributing to any improvement in participation outcomes. Using the Emotiv EPOC$^+$ headset (previous iteration of Emotiv EPOC$^\textrm{x}$), \citet{zhang2019evaluating} determined that typically developing children between the ages of 6 and 18 "can quickly achieve control and execute multiple tasks using simple EEG-based BCI systems", supporting the introduction of this instrumentation to children with disabilities and/or complex medical needs. Considering this evidence, and that readily available hardware and software is more likely to reach a larger number of potential users, the Emotiv EPOC$^\textrm{x}$ headset was chosen for the BCI clinic since it is a commercially available, wireless BCI headset based on electroencephalography (EEG) technology. With 14 saline-based electrodes that eliminate the inconvenience of electrode gel, the headset is easy to set up and is therefore conducive to in-home use and clinical applications. 

Another essential component of a pediatric BCI solution is a software system that supports interactions between the hardware, the user interface, and other applications. The Emotiv system supports toolboxes such as NodeRed that allow users to interface Emotiv with external applications for BCI integration. However, for users who do not know how to program, it may be challenging to create and maintain such an interface.

In this paper, we present EmoconLite\footnote{\url{http://hollandbloorview.ca/emocon}} - a freely available BCI software system developed for the Holland Bloorview BCI clinic that can be used with the Emotiv hardware. The EmoconLite application provides a simple interface that allows users to access media. It also allows users to play BCI-compatible games, both locally and online. In section \ref{architecture}, we explain EmoconLite's software architecture and functionality. We discuss the role of EmoconLite within the Holland Bloorview BCI clinic in section \ref{discussion}, and conclude our paper in section \ref{conclusion}.

\section{Application Architecture}
\label{architecture}
\begin{figure}
\centering
\scalebox{.7}{
	\begin{tikzpicture}[node distance=1.6cm ]
	\node (pro1) [process] {User};
	\node (pro2) [process, below of=pro1] {EmoconLite};
	\node (pro3) [process, below of=pro2] {Cortex API};
	\node (pro4) [process, below of=pro3] {Emotiv headset};
	\node (pro5) [process, right of=pro2, xshift=3cm] {Activities
	    \begin{itemize}
	        \item YouTube
	        \item HelpKidzLearn
	        \item Brain Joust
	        \item Steam
	    \end{itemize}
	};
	
	\draw [arrow] (pro1) -- (pro2);
	\draw [arrow] (pro2) -- (pro5);
	\draw [arrow2] (pro2) -- (pro3);
    \draw [arrow2] (pro3) -- (pro4);
    
	\end{tikzpicture}
}
\caption{EmoconLite architecture}
\label{EmoconLiteArch}
\end{figure}
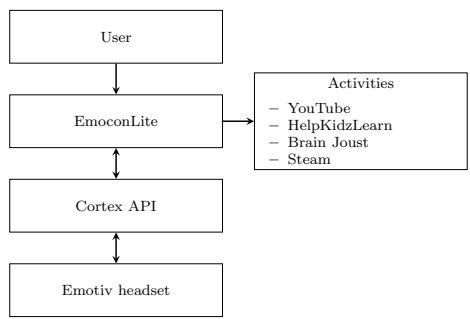

EmoconLite is a software application designed to bridge the gap between the EmotivBCI system and other computer programs (see Figure \ref{EmoconLiteArch}). The application is implemented in \mbox{Python 3.7} \citep{python} using PySide2 \citep{pyside}. The app is cross-platform and can be run on Windows, Linux and macOS with small adjustments. The EmoconLite graphical user interface (GUI) is divided into five sections based on the user workflow: \emph{Setup and training}, \emph{Activity}, \emph{Sensitivity}, \emph{Controls}, and \emph{Status}.

\subsection{User Workflow}
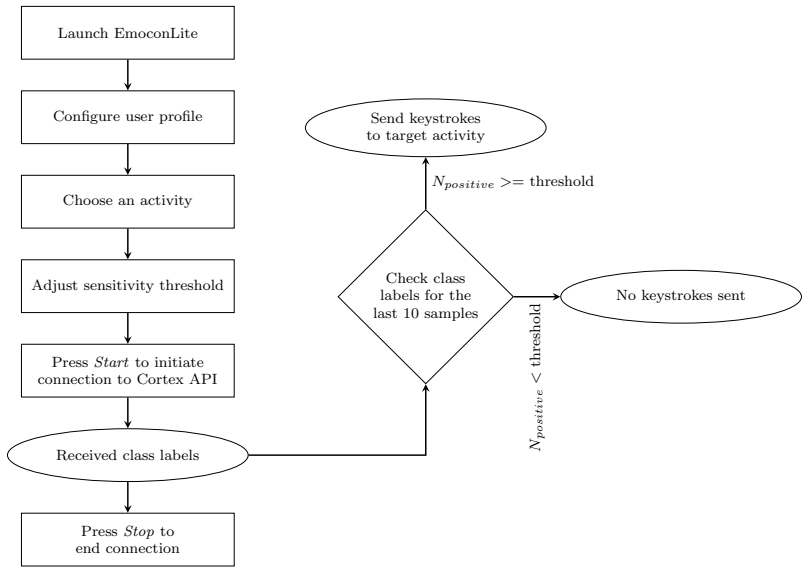
\begin{figure}
\centering
\scalebox{.7}{
	\begin{tikzpicture}[node distance=1.6cm ]
	\node (pro1) [process] {Launch EmoconLite};
	\node (pro2) [process, below of=pro1] {Configure user profile};
	\node (pro3) [process, below of=pro2] {Choose an activity};
	\node (pro4) [process, below of=pro3] {Adjust sensitivity threshold};
	\node (pro5) [process, below of=pro4] {Press \emph{Start} to initiate connection to Cortex API};
    \node (io1) [io, below of=pro5] {Received class labels};
    \node (dec1) [decision, right of=io1, xshift=4cm, yshift=3cm] {Check class labels for the last 10 samples};
    \node (io2) [io, above of=dec1, yshift=1.6cm] {Send keystrokes to target activity};
    \node (io3) [io, right of=dec1, xshift=3.2cm] {No keystrokes sent};
    \node (pro6) [process, below of=io1] {Press \emph{Stop} to end connection};

	\draw [arrow] (pro1) -- (pro2);
	\draw [arrow] (pro2) -- (pro3);
    \draw [arrow] (pro3) -- (pro4);
    \draw [arrow] (pro4) -- (pro5);
    \draw [arrow] (pro5) -- (io1);
    \draw [arrow] (io1) -| (dec1);
    \draw [arrow] (dec1) -- (io2) node[midway,sloped,right,rotate=270]{$N_{positive} >= $ threshold};
    \draw [arrow] (dec1) -- (io3) node[midway,sloped,left,rotate=90]{$N_{positive} < $ threshold};
    \draw [arrow] (io1) -- (pro6);
	\end{tikzpicture}
}
\caption{User workflow diagram, where $N_{positive}$ is the number of positive labels, and \emph{threshold} is the chosen value for the sensitivity}
\label{workflow}
\end{figure}

The workflow of EmoconLite is depicted in Figure \ref{workflow}.

\subsubsection{Setup and training}
The \emph{Setup and training} section allows users to train the EmotivBCI system \citep{emotivBCI} to recognize different mental states. In addition to the neutral mental state, there are 13 mental tasks users can configure, allowing up to 13 different inputs to be provided. The current version of EmoconLite accepts binary tasks (e.g. neutral vs any mental task), and translates them into "on" and "off" commands. However, a future iteration of the application will handle multiple tasks and map them to arrow keys to provide more degrees of control. In this section, users can create a new profile, train the BCI model to recognize their chosen mental tasks, and configure their profile to link mental tasks to different commands.

\subsubsection{Activity}
In the \emph{Activity} section, the configured profiles can be used for a list of four activities: to launch a YouTube video clip and toggle the play and pause functions, to play HelpKidzLearn games\footnote{\url{http://https://www.helpkidzlearn.com/}}, to play the Brain Joust game\footnote{\url{https://www.bci.games/games/brain-joust}}, or to play Steam games\footnote{\url{https://store.steampowered.com/}}. Clicking on the YouTube button will open up a new browser window, where users can select their desired YouTube video clip. The user can then perform a mental task (i.e. any task other than neutral) to play or pause the video clip. Users can also access HelpKidsLearn, an online learning platform for children of all abilities, which consists of accessible games and activities. Another activity available to users is Brain Joust - an accessible video game developed by BCI Games\footnote{\url{https://www.bci.games/}}. Finally, EmoconLite also supports a variety of games available on the video game distribution service, Steam. By leveraging Steam's Remote Play functionality, users can play online and multiplayer games with their friends over the internet.

\subsubsection{Sensitivity}
Users can toggle an interactive slider to adjust the sensitivity threshold of EmoconLite's classification algorithm. This algorithm uses Emotiv's internal classifier model to classify EEG samples as neutral or as one of the configured mental tasks. The predicted label of a given sample is determined based on the classification of a window of 10 samples. In turn, the predicted output at a given point in time is determined based on the evidence accumulated from the 10 most recent classifications, corresponding to approximately the previous second of the EEG signal. In other words, if the number of samples labeled as a particular class within this preceding period exceeds the sensitivity threshold, the current sample is assigned the corresponding label. Users can increase the sensitivity threshold if there are too many false activations and decrease the sensitivity threshold if there are too many missed activations.

\subsubsection{Controls}
When the user wants to use EmoconLite to send commands to their chosen activity, their caregiver can press the \emph{Start} button. EmoconLite will then initiate a background thread to communicate with the Emotiv headset through a WebSocket using the Emotiv Application Programming Interface (API) called Cortex API\footnote{\url{https://emotiv.gitbook.io/cortex-api/}}. After making sure the headset is connected and the proper user profile is selected, the thread will start a session to receive a stream of classification labels from the Cortex API and simulate and send a keystroke press to the selected application. When the user is done with the activity, the caregiver can press the \emph{Stop} button to disconnect the Emotiv headset and to stop it from providing further input.

\subsubsection{Status bar}
All the messages related to the initiation and execution of commands are shown in the status bar. If any error occurs while connecting to the Emotiv headset and Cortex API, a corresponding message will appear in the status bar to help users troubleshoot the problem. During an activity, the number of positive (i.e. mental task) and negative (i.e. neutral) class labels over the last 10 samples will also be shown in the status bar. This makes it easier for the user to adjust the sensitivity threshold.

\subsection{Data structure}
Data streamed to and from the Cortex API are in are in JavaScript Object Notation (JSON), an open standard file format. The user's mental tasks can be accessed by subscribing to the Cortex API. EmoconLite continuously parses the JSON output object from the Cortex API and checks if the desired mental task has been performed or not, based on the model created for each user internally in the EmotivBCI application. The appropriate keystroke simulation will be sent to the target activity based on this output.

\subsection{Screenshot}
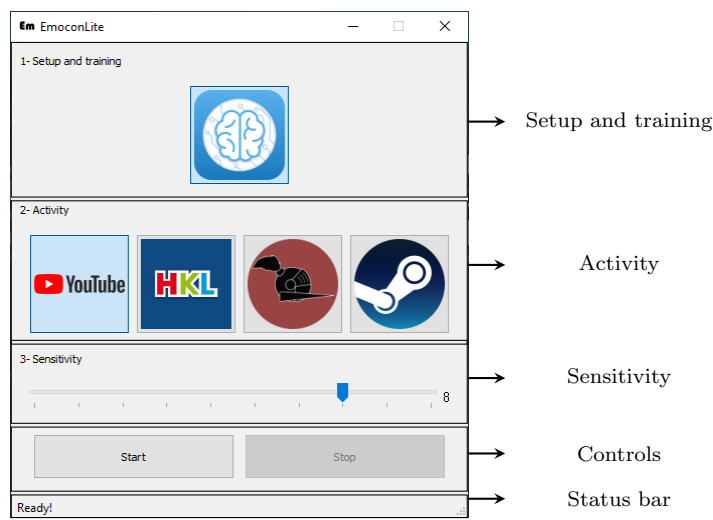
\begin{figure}
    \begin{tikzpicture}
        \draw (0, 0) node[inner sep=0] {\includegraphics[scale=.5]{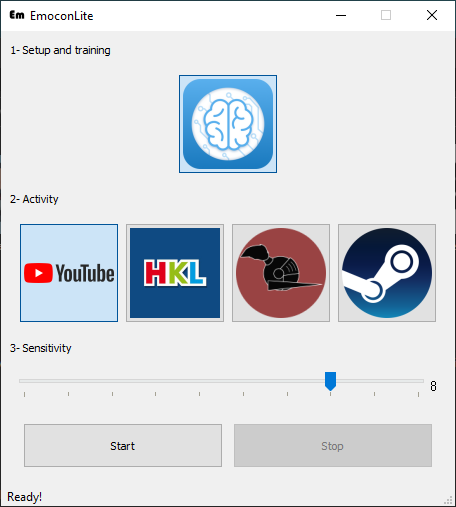}};
        \draw[thin] (-3,.9) rectangle (3,2.95);
        \draw[arrow] (3,1.9) -- (3.5,1.9);
        \draw (5,1.9) node {Setup and training};
        
        \draw[thin] (-3,-1.05) rectangle (3,.85);
        \draw[arrow] (3,0) -- (3.5,0);
        \draw (5,0) node {Activity};
        
        \draw[thin] (-3,-2.1) rectangle (3,-1);
        \draw[arrow] (3,-1.5) -- (3.5,-1.5);
        \draw (5,-1.5) node {Sensitivity};
        
        \draw[thin] (-3,-3) rectangle (3,-2.15);
        \draw[arrow] (3,-2.5) -- (3.5,-2.5);
        \draw (5,-2.5) node {Controls};
        
        \draw[thin] (-3,-3.35) rectangle (3,-3.05);
        \draw[arrow] (3,-3.1) -- (3.5,-3.1);
        \draw (5,-3.1) node {Status bar};
    \end{tikzpicture}
\caption{GUI of the EmoconLite app}
\label{EmoconLite}
\end{figure}

Figure \ref{EmoconLite} shows the GUI of EmoconLite. The GUI layout intuitively follows each step of the user workflow.

\subsection{Documentation}
A quick guide documentation\footnote{\url{http://hollandbloorview.ca/emocon}} has been created to guide users through the step-by-step process of downloading and setting up EmoconLite, along with the supported activities such as Brain Joust and Steam. This setup prepares EmoconLite to communicate with EmotivBCI on the computer. The quick guide is an essential reference for family and clinicians to provide set up and training for children. It is made available in Dutch, English, French, Chinese, Persian, Portuguese, to reduce language barriers.

\section{Discussion}
\label{discussion}
Play is the primary occupation of children \citep{hoogsteen2010can,lynch2016play}. This instinctive ability and human right to play allows for childhood discovery by observation and exploration of the world around them, supporting growth, learning, and development. Active, engaged and independent play participation within leisure and recreation experiences has many benefits to childhood development \citep{mcnaughton2013ipad,larson1999children,crawford2014strategies,king2014integrated,longo2014cross,chiarello2017excellence,chien2017parent}. Based on clinical observation, evidence-based and evidence-informed practice, there is an identified need to support children with severe disabilities and/or medical complexities in their active and independent engagement in a therapeutic recreation (TR) BCI program via assisted technology \citep{ramella2009making,deavours1994lifespace}. Rather than focusing on switches that require physical movement that may remain elusive, a BCI focuses on its user's internal capacity, utilizing their brain signals to mobilize their capabilities to engage and self-initiate participation actively. BCI, as an access method, activates the users' strengths and abilities, to enhance capabilities \citep{hood2016strengths,wise2018integrating}.

\citet{wise2018integrating} described capability as "the freedom to choose from a set of opportunities related to what one wants to do and be". Specifically, real freedom is to pursue, achieve and participate in activities of life that are deeply valued \citep{wise2018integrating}. A capability- and strength-based approach is imperative to support the human dignity of children with disabilities and is essential to their well-being \citep{sayer2011things,mietola2017voiceless,hood2016strengths,wise2018integrating}. A BCI approach requires co-creation and collaboration between clients, families, clinicians, and their supporting partners to support dignity, capabilities, health, and the well-being of children with disabilities \citep{mumford2014access,mumford2016application}. All of these stakeholders will require access to the technology, training with the technology, co-creation of participatory experience, and participatory action research and action sensitive research within a psychologically safe environment.

Utilizing the response efficiency theory, \citet{mumford2014access}, created a collaborative, child and family-centred approach to support the implementation of access technology for children with severe and multiple disabilities called the Access Technology Delivery Protocol (ATDP). Addressing the need for standardization of assistive technology  prescription, a process that "...has been described as an intrinsically difficult procedure that is prone to failure" with a 30\% rate of device abandonment \citep{mumford2014access,mumford2016application}. The ATDP when implemented within the BCI Clinic and combined with the TR BCI program may support  clients and families to sustain use of the technology through fun, function and social participation. 

\citet{huber2011should} proposed that health is the "ability to adapt and self manage in the face of social, physical, and emotional challenges". Children with severe disabilities and/or medical complexities require effective and standardized approaches to assistive technology prescription that provide adaptations to support active, engaged and independent participation within leisure and life which may result in improved health benefits, well-being and quality of life \citep{mumford2014access,mumford2016application}. 

Adaptations that consider BCI may support health and well-being, reducing leisure barriers in the short term, build leisure competence, and build skills for future communication applications in the long term. The EmoconLite software and Emotiv headset are assistive technology adaptations that may support this aim when implemented with the ATDP process \citep{mumford2016application}. At Holland Bloorview Kids Rehabilitation Hospital and partnering BCI clinics across Canada, children with disabilities, along with their families and clinicians have access to this adaptation strategy through the BCI clinic \citep{kinney2020advancing}. This will support childhood development through play and leisure pursuits both independently and with peers in a social environment to reduce leisure barriers and increase active, independent and engaged play participation.

The Emotiv headset paired with the EmoconLite creates a new access method to engage in leisure and recreation pursuits, which can help empower choice, control and self-initiated engagement by the user \citep{king2014integrated}. These are essential building blocks for intrinsic feelings of autonomy and agency for the growing and developing child, which influence intrinsic motivation \citep{steinert2019doing}. During play, the results of the activation of the effector device, such as seeing a character move in a video game or hearing their favourite music play within a YouTube video, influences extrinsic motivation to participate. Intrinsic and extrinsic motivation through play-based experience may support the sustainable use of BCI technology, reducing device abandonment.

Through interactions with the environment, as well as communication with friends, family and the community, children learn, grow, share, empathize and understand \citep{wise2018integrating}. This kind of agentic engagement sets a foundation for learning \citep{montenegro2017understanding}. When a child with severe disabilities and/or medical complexities is supported through adaptations to have an opportunity to create an action with a particular effect, their world will open up with growth \citep{larson1999children,wehmeyer2005self}. Future research that considers the Emotiv headset paired with the EmoconLite as an adaptation to support childhood development through self-determined and agentic experiences may further advance our understanding of the benefits of this proposed approach.

Social opportunities are a necessity for growth, learning and development; and ultimately health and well-being. \citet{sexton2015overlooked} looked at social factors to improve BCI's effectiveness and noted that "BCIs are not just about brains but also about brains in interaction with other brains". Humans are inherently social creatures that require interaction, reciprocity, communication partners to effectively function, learn, grow, develop and have a meaningful life. Therefore, aside from the much needed interdisciplinary training programs that utilize ATDP, to determine whether this technology adaptation is a good fit, the BCI clinic will also focus on individualized co-created goals and life skills within leisure and recreation \citep{mumford2014access,mumford2016application,jessup2007interdisciplinary,pritchard2020child,king2013four,deavours1994lifespace,ramella2009making}. Social opportunities, facilitated within TR programs, will support clients in practising and having fun while aiming to reach their goals, and encourage the sustained use of the technology.

With the creation of the EmoconLite application, Emotiv technology becomes more practically accessible. In its application within the BCI clinic, EmoconLite can better support play and skill development for children with severe disabilities and/or medical complexities during leisure. To the best of our knowledge, this is the first effort to design, implement and provide a freely available application to the users who are interested in utilizing Emotiv headsets to perform a task within an interdisciplinary training environment and a TR BCI program. 

Due to the world's current status, the pandemic has created unique challenges requiring creative solutions to healthcare. As the world has moved to online formats, such as Zoom and Ontario Telemedicine Network, so has the BCI clinic. A BCI-at-home virtual program is the devised strategy across the BCI-CAN Network to support client and family continued access and engagement during the current COVID-19 pandemic \citep{kinney2020advancing}. One of the BCI clinic's aims is to ensure the integration of the BCI systems within the community, especially at home where experiences can be shared with family and friends. These experiences can also encourage the user, further motivating continued engagement.

One of the notable limitations of EmoconLite is the number of mental tasks it supports. Since the current version of the software only supports binary classification to distinguish between the neutral state and mental tasks, this limits the number of inputs users can provide to external applications. Another limitation of EmoconLite is that it builds upon the internal classification algorithm of the EmotivBCI software. Since this is a proprietary algorithm, the lack of transparency makes it difficult to understand the underlying classification approach or to potentially extend it. Future iterations of the application would expand the support for more mental tasks to give users more degrees of control and implement open-source classification algorithms to allow developers to improve on the classification performance. It would also be beneficial to expand the repository of supported activities and incorporate accessibility features in the GUI of the EmoconLite application. In addition, future works will also look into extending EmoconLite to control a variety of effector devices, such as Bluetooth toys and devices. This will make EmoconLite even more functional and versatile.

BCI systems have a lot of potential as an alternative access method and as an augmentative and alternative communication method for children with severe disabilities and/or medical complexities \citep{kinney2020advancing,brumberg2018brain}. EmoconLite will support leisure experience for children with severe disabilities and/or medical complexities, giving them the opportunity to choose to initiate an activity and engage actively and independently through assistive technology. As such, continued development of BCIs for recreation, leisure and life pursuits will be essential in the well-being of these children. Through co-created participatory action research and action sensitive research, the interdisciplinary training program and the TR BCI Program at Holland Bloorview Kids Rehabilitation Hospital will support further iterations of the development of EmoconLite to meet client and family needs. 

Additionally, communication is a huge aspect of social experience as well as co-created leisure and recreation experience. Leisure interests, preferences, and choice can be determined through connectivity and communication. These are essential components for participation, the development of leisure and life goals, and agency. Children acquiring BCI skills through leisure and play will become trained in BCI, and these skills may be applicable to potential communication-based BCIs in the future. Development of the hardware and software, considerations of how BCIs can support recreation and leisure engagement, environmental controls, and communication are all areas that require co-created exploration to inform user-centred design.

\section{Conclusion}
\label{conclusion}
This paper introduced a new freely available software application called EmoconLite, which is an end-to-end solution that brings BCI applications to users with no programming experience. Building upon the Emotiv hardware and software system, users can control external computer programs such as YouTube, HelpKidzLearn, Brain Joust, and Steam in real-time using their EEG signals. When integrated into Holland Bloorview’s BCI clinic, EmoconLite can support children with disabilities and/or medical complexities by enabling access to BCI technologies to help them adapt, benefiting their well-being and improving their quality of life. Strategies on maintaining, motivating, mitigating frustration, and reducing device abandonment while meeting clients' needs to develop, learn and grow will need to be considered along with continued collaboration across the board.

In the future, we will be extending the support for recognizing more mental tasks, while expanding the repository of games and effector devices, connecting to Bluetooth toys and devices to provide extra functionalities to the users. These improvements will be followed by making the next iteration of EmoconLite target-agnostic to control various applications. By expanding functionalities, a variety of activities will become available, opening up additional avenues for exploration. Consultation with and inclusion of clients, families and clinicians involved in the BCI programs will inform new iterations of EmoconLite to meet client interests and needs. Through the BCI clinic, play and leisure is transformed. Children with severe disabilities and/or medical complexities, and their families will have an opportunity to adapt and cope with their congenital and acquired conditions through their brain signals, and explore their potential, leading to resilience in the face of challenges while reducing barriers to leisure participation and supporting leisure values and leisure competence.

\begin{acknowledgements}
This work was supported partially by Mitacs through the Mitacs Elevate program and Holland Bloorview Kids Rehabilitation Hospital Foundation. We want to thank Sarah House and Leslie Mumford at PRISM Lab, Heather Keating, Stephanie Hicks, Rachel Arsenault and Kendra Abbey at Holland Bloorview Kids Rehabilitation Hospital, Dr. Maureen Connolly at Brock University, all members of the BCI-CAN network for their constructive feedback and suggestions and those who supported the translation of the EmoconLite user guides; Wendy Cox, Danielle Miranda, and Jenny Tou.
\end{acknowledgements}

\section*{Conflict of interest}
The authors declare that they have no conflict of interest.

\bibliographystyle{abbrvnat}
\bibliography{mybib}

\end{document}